\documentclass{aipproc}

\layoutstyle{8x11single}

\usepackage{amsmath,epsfig}
\usepackage{amssymb,amsfonts}
\usepackage{latexsym}

\relax

\usepackage{amsmath}
\usepackage{amssymb}
\usepackage{graphicx}

\newcommand{\AP}{\alpha^{\prime}}
\newcommand{\pd}{\partial}

\newcommand{\Fc}{\mathcal{F}}
\newcommand{\Gc}{\mathcal{G}}

\begin{document}

\title[SFT non-locality in cosmology]{SFT non-locality in cosmology: solutions, perturbations and observational evidences}

\author{Alexey~S.~Koshelev\footnote{Postdoctoral researcher of FWO-Vlaanderen, e-mail:~\texttt{alexey.koshelev@vub.ac.be}.}}{
 address={Theoretische Natuurkunde, Vrije Universiteit Brussel and
The International Solvay Institutes, Pleinlaan 2, B-1050 Brussels, Belgium}}


\begin{abstract}
In this note cosmological models coming out of the String Field Theory (SFT) in application to the Dark Energy are reviewed.
A way of constructing solutions in the case of linear models is outlined, cosmological perturbations and observational evidences of such models
are explored. We explicitly demonstrate the stability of the system at the linear order in the most typical configuration.
\end{abstract}



\maketitle


\section{Introduction}

In this note we briefly review the recent activity on a new class of cosmological models based on the string field
theory (SFT) (for details see reviews~\cite{review-sft}) and the
$p$-adic string theory \cite{padic} focusing on their applications to the present Dark Energy phenomenon.
The reader is referred to \cite{IA1,Koshelev07,AKd,KVlast} and refs. therein for more detailed analysis on the subject.
It is known that the SFT and the $p$-adic string theory are
UV-complete ones. Thus one can expect that resulting (effective) models
should be free of pathologies. Furthermore, models originating from the
SFT exhibit one general non-standard property, namely, they have terms
with infinitely many derivatives, i.e. non-local terms. The higher
derivative terms usually produce the well known Ostrogradski
instability \cite{ostrogradski}.
However the Ostrogradski result is related to higher than two but
finite number of derivatives. In the case of infinitely many
derivatives it is possible that instabilities do not appear.

Contemporary cosmological
observational data~\cite{data,Komatsu} strongly support that the
present Universe exhibits an accelerated expansion providing thereby an
evidence for a dominating DE component~\cite{review-de}. Recent results of WMAP~\cite{Komatsu} together
with the data on Ia supernovae give the following bounds for the DE
state parameter $ w_{\text{DE}}=-1.02^{+0.14}_{-0.16} $. Note that the
present cosmological observations do not exclude an evolving DE state
parameter $w$.
Non-local models of the type one has from the SFT may have effective phantom behavior and are of
interest for the present cosmology. To construct a stable model with
$w<-1$ one should construct the effective theory with the Null Energy Condition (NEC) violation
from the fundamental theory, which is stable and admits quantization.
This is a hint towards
the SFT inspired cosmological models.

In this note we shall demonstrate how an effective phantom behavior may appear, shall consider an indicative example along with
cosmological perturbations arguing that the phantom phase in these models does not produce growing perturbations and shall
outline a possible signature of such non-local models.


\section{The model}

We are interested in non-local models arising from SFT in the far asymptotic regime for large times. The starting point is the Lagrangian
\begin{equation}
S=\int d^4x\sqrt{-g}\left(\frac{R}{16\pi
G_N}+\frac1{g_o^2}\left(-\frac12\pd_\mu T\pd^\mu
T+\frac1{2\AP}T^2-\frac1{\AP}{v(\bar T)}\right)-\Lambda'\right).
\label{action_model_pre}
\end{equation}
where we manifestly indicate four dimensions despite the stringy origin. We work in $1+3$ dimensions with the signature $(-,+,+,+)$, the coordinates are denoted by
Greek indices $\mu,\nu,\dots$ running from 0 to 3. Spatial indexes are $a,b,\dots$ and they run from 1 to 3. Such four-dimensional action is motivated by the string field theory and details can be found in \cite{Koshelev07}.
$G_N$ is the Newtonian constant: $8\pi G_N=1/M_P^2$, where
$M_P$ is the Planck mass, $\AP$ is the string length squared (we do not
assume $\AP=1/M_P^2$), $g_o$ is the open string coupling constant. $g_{\mu\nu}$ is the metric tensor, $R$
is the scalar curvature, $\Lambda'$ is a constant, $\bar T=\Gc(\AP\Box)T$ with
\begin{equation}
 \Box=D^{\mu}\pd_{\mu}=\frac1{\sqrt{-g}}\partial_{\mu}\sqrt{-g}g^{\mu\nu}\partial_{\nu}
\end{equation}
 and $D_\mu$ being a covariant
derivative, $T$ is a scalar field primarily associated with the open
string tachyon. The function $\Gc(\AP\Box)$ may be not a polynomial
manifestly producing thereby the
non-locality. In SFT one has $\Gc(\AP\Box)=e^{-\frac{\beta}2\AP\Box}$ but we keep this function general.
$\beta$ is a parameter determined exclusively by the conformal field theory of the string. Here we see the very important feature: we may have non-locality in the action. We stress here that appearance of non-localities is general feature of the SFT based models and exactly the one we are going to explore\footnote{Appearance of higher derivatives is not unique for this theory. Non-commutative theories, for instance, also have higher derivative, but these non-local structures are very different.}.
Fields are
dimensionless while $[g_o]=\text{length}$. $v(\bar T)$ is an open
string tachyon self-interaction. It does not have a quadratic in $T$ term. Factor $1/\AP$ in front of the tachyon
potential looks unusual and can be easily removed by a rescaling of
fields. For our purposes it is convenient keeping all the fields dimensionless.

Redefining the tachyon field $T_b=\bar T$ and introducing dimensionless coordinates
$\bar{x}_\mu=x_\mu/\sqrt{\AP}$ the
above action can be rewritten as (where we omit bars for simplicity and put hereafter $\AP=1$)
\begin{equation}
S=\int d^4x\sqrt{-g}\left(\frac{R}{16\pi
G_N}+\frac1{g_o^2}\left(-\frac12\pd_\mu \tilde T_b\pd^\mu
\tilde T_b+\frac1{2}\tilde T_b^2-{v(T_b)}\right)-\Lambda'\right). \label{action_model}
\end{equation}

Let us emphasize that the potential of the field $T_b$ is $V=-\frac1{2}T_b^2+v(T_b)$. Given an extremum of the potential $V$ one can linearize the theory around it using $T_b=T_0-\tau$. This gives $V=-\frac12\tau^2+\frac{v(T_0)''}2\tau^2+V(T_0)$.
Action (\ref{action_model}) being linearized around the extremum of the potential of the field $T_b$ can be written as
\begin{equation}
S=\int d^4x\sqrt{-g}\left(\frac{R}{16\pi
G_N}+\frac1{2g_o^2}\tau\Fc(\Box)\tau-\Lambda\right),
\label{action_model2}
\end{equation}
where
\begin{equation}
    \Fc(\Box)=(\Box+1)\Gc^{-2}(\Box)-m^2.
\end{equation}
Here $m^2\equiv v(T_0)''$ and $\Lambda$ accounts $\frac{V(T_0)}{g_o^2}$. From SFT one has
\begin{equation}
    \Fc_{\text{SFT}}(\Box)=(\Box+1)e^{\beta\Box}-m^2
\end{equation}
$\Fc$ is in fact the inverse propagator and it is natural to expect $\beta<0$ in the SFT case corresponding to a convergent propagator at large momenta. In this SFT example we can learn a very important thing. Assume $m=0$. Then $\Fc_{\text{SFT}}=(\Box+1)e^{\beta\Box}$ and the non-locality does not show up at all. Indeed, the poles of the propagator do not feel the exponent. Other way of thinking is that the exponential factor can be hidden by a field redefinition. The situation is dramatically different for $m\neq0$. The propagator $1/\Fc_{\text{SFT}}$ has infinitely many poles. This is the manifestation of the non-locality. Also we stress that physics would be totally different for full function $\Fc$ and its truncated series expansion because the pole structure may get modified considerably.

We see that the most important role is played by the spectrum of the theory which is determined by the equation
\begin{equation}
    \Fc(J)=(J+1)\Gc^{-2}(J)-m^2=0.\label{cheq}
\end{equation}
We name it \textit{characteristic} equation. It can be an algebraic or a transcendental one. Thus until examples are considered we shall keep $\Fc(\Box)$ general. The only assumption is that all roots are simple. Also the analyticity of the function $\Fc$ on the complex plane is important so that one can represent $\Fc$ by the convergent series
expansion:
\begin{equation}
\Fc=\sum\limits_{n=0}^{\infty}f_n\Box^n\text{ and }f_n\in\mathbb{R}.
\end{equation}
Reality of coefficients is required by the hermiticity of the Lagrangian.
Strictly speaking even the analyticity requirement can be weakened but in this case one has to be careful with poles and consider only the appropriate domains of the argument were the series expansion does converge.

Equations of motion are
\begin{eqnarray}
G_{\mu\nu}=\frac{8\pi G_N}{g_o^2}T_{\mu\nu}&=&{4\pi G}\sum_{n=1}^\infty
f_n\sum_{l=0}^{n-1}\left(\pd_\mu\Box^l\tau\pd_\nu\Box^{n-1-l}\tau
+\pd_\nu\Box^l\tau\pd_\mu\Box^{n-1-l}\tau-\right.\nonumber\\
&&\left.-g_{\mu\nu}\left(g^{\rho\sigma}
\pd_\rho\Box^l\tau\pd_\sigma\Box^{n-1-l}\tau+\Box^l\tau\Box^{n-l}\tau\right)\right)-8\pi G_Ng_{\mu\nu}\Lambda,
\label{EOM_g}\\
\Fc(\Box)\tau&=&0
\label{EOM_tau}
\end{eqnarray}
where $G_{\mu\nu}$ is the Einstein tensor, $T_{\mu\nu}$ is the
energy-momentum (stress) tensor and $G\equiv G_N/g_o^2$ is a
dimensionless analog of the Newtonian constant. It is easy to check
that the Bianchi identity is satisfied on-shell and for $\Fc=f_1\Box+f_0$ the usual energy-momentum tensor for
the massive scalar field is reproduced. Note that equation
(\ref{EOM_tau}) is an independent equation consistent with system
(\ref{EOM_g}) due to the Bianchi identity.

Classical solutions to equations (\ref{EOM_g}) and (\ref{EOM_tau}) were
studied and analyzed in the literature.
The cornerstone in this analysis is the fact that action (\ref{action_model2}) is fully equivalent to the following action with many free local scalar fields
\begin{equation}
S_{local}=\int d^4x\sqrt{-g}\left(\frac{ R}{16\pi
G_N}-\frac{1}{g_o^2}\sum_i\frac{\Fc'({J}_i)}{2}\left(g^{\mu\nu}\pd_\mu\tau_i\pd_\nu\tau_i
+{J}_i\tau_i^2\right)-\Lambda\right) \label{action_model_local}
\end{equation}
Here there are as many scalar fields as many roots the characteristic equation (\ref{cheq}) has.

To see the equivalence first we write down the equations of motion for the latter local action which are
\begin{equation}
\begin{split}
G_{\mu\nu}=\frac{8\pi G_N}{g_o^2}T_{\mu\nu}&=\frac{8\pi G_N}{g_o^2}\sum_i\Fc'({J}_i)\left(\pd_\mu\tau_i\pd_\nu\tau_i
-\frac1{2}g_{\mu\nu}\left(g^{\rho\sigma}\pd_\rho\tau_i\pd_\sigma\tau_i+{J}_i\tau_i^2\right)
\right)-\\
&-{8\pi G_N}g_{\mu\nu}\Lambda 
\end{split}\label{EOM_g_onshell}
\end{equation}
and
\begin{equation}
\Box\tau_i=J_i\tau_i,~\text{for all}~i\label{EOM_taui_onshell}
\end{equation}
Note that $J_i$ is a root of characteristic equation (\ref{cheq}). Therefore any $\tau_i$ solves equation (\ref{EOM_tau}) just because
\begin{equation*}
\Fc(\Box)\tau_i=\Fc(J_i)\tau_i=0,~\text{for all}~i
\end{equation*}
As a consequence of linearity of equation (\ref{EOM_tau}) its solution is a linear combination $\sum_i\tau_i$.
Any arbitrary coefficients in this summation can be accounted by the integration constants in each $\tau_i$. Substituting $\tau=\sum_i\tau_i$ in (\ref{EOM_g}) one can readily find that equation (\ref{EOM_g}) is transformed into equation (\ref{EOM_g_onshell}). To complete the proof of the equivalence statement one has to show that $\tau=\sum_i\tau_i$ is a general solution to equation (\ref{EOM_tau}). This was done for a large class of functions $\Fc(\Box)$.

It is important to understand at this point that roots $J_i$ as well as coefficients $\Fc'(J_i)$ can be complex.
This should not spoil the theory just because all the local scalar fields are just mathematical functions.
They do not have physical meaning. What is important is that the original field $\tau$ must be real since
it represents a physical excitation. The latter is easy to achieve. Namely, the roots $J_i$ are either real
or complex conjugate. Complex conjugate $J_i$ would yield complex conjugate (up to an overall constant factor)
solutions for $\tau_i$. Thus, it is a matter of choice of the integration constants to make the linear combination
of all $\tau_i$ real. Moreover, coefficients $\Fc'(J_i)$ can be absorbed in the scalar fields by the field rescaling.
We do not do this just because real $\Fc'(J_i)$ would indicate whether the corresponding field is a phantom or not.
Also we note that the above proven equivalence is true for arbitrary metric.

We continue with local action (\ref{action_model_local}). Moreover, from now on we specialize to the spatially flat Friedmann--Robertson--Walker (FRW) Universe.
The  metric is of the form
\begin{equation}
\label{mFr} ds^2={}-dt^2+a^2(t)\left(dx_1^2+dx_2^2+dx_3^2\right)
\end{equation}
where $a(t)$ is the scale factor, $t$ is the cosmic time. The Hubble parameter is as usual $H=\dot a/a$ and the dot hereafter in this paper denotes a derivative with respect to the cosmic time $t$.
Background solutions for $\tau$ are taken to be space-homogeneous
as well.
Equations of motion become much simpler.
\begin{equation}
\label{FrEOMgFRW}
\begin{split}
3H^2&=4\pi G\sum_i\Fc_{,{M}}'({M}_i)\left(\dot\tau_i^2
+{M}_i\tau_i^2\right)+8\pi G_N\Lambda,~
\dot H=-4\pi G
\sum_i\Fc_{,{M}}'({M}_i)\dot\tau_i^2.\\
\end{split}
\end{equation}
and
\begin{equation}
\ddot\tau_i+3H\dot\tau_i+J_i\tau_i=0,~\text{for all}~i
\label{FrEOMtauFRW}
\end{equation}

It was proven in \cite{KVlast} that cosmological perturbations are also equivalent in the free theory with one
non-local scalar field (\ref{action_model2}) and in the corresponding local theory (\ref{action_model_local}) with
many scalar field. 
In the next Section we shall consider the most important example with two complex conjugate roots $J_i$.

Obviously if we have two consequent real roots $J_1$ and $J_2$ then $\Fc'(J_1)$ and $\Fc'(J_2)$ are real and
would have different signs. This is one the origin of the phantom like behavior in such models. Such phantom behavior,
however, is not distinguishable from the already known examples. Complex roots $J_i$ in turn also produce the phantom-like behavior
but this one is very different from the previous case. Indeed, suppose we have only two complex conjugate roots $J_1$ and $J_2=J_1^*$ in action
(\ref{action_model_local}). Then one can easily pass to the real components of the fields. However, such real fields will be quadratically coupled.
Thus, the intuition based on the signs of kinetic terms and mass terms will be broken. New behavior is reflected in the fact that
solutions are classically stable and perturbations do not grow even though the phantom divide can be crossed.


\section{Example of background solution}

The reduction to a one-field model is possible but practically is trivial. Nothing new arises because 
the structure coming from the SFT is not felt. One pair of complex conjugate roots $J_1=J$ and $J_2=J^*$ is a new situation
worth the attention.

It is obvious that if a complex number $J$ is a root of $\Fc$, then
$J^*$ is a root of $\Fc$ as well.  The system of Friedmann equations becomes:
\begin{equation}
\label{FrEOMtauc}
\begin{split}
3H^2&=4\pi G\left[\Fc'(J)\left(\dot\tau_1^2
+J\tau_1^2\right)+{\Fc'({J^*})}\left(\dot{\tau_2}^2
+{J^*}{\tau_2}^2\right)\right]+8\pi G_N\Lambda,\\
\dot H&=-4\pi G
\left[\Fc'(J)\dot{\tau_1}^2+\Fc'({J^*})\dot{\tau_2}^2\right].\\
\end{split}
\end{equation}
This is a really new configuration much less explored in the past. The
distinguishing feature is the complex coefficients. Because of this
solutions for the scalar fields also expected to be complex. However,
as it was already stressed these fields are not physical. What is
physical is their linear combination and only the latter must be real.
This is easy to achieve adjusting the integration constants, say,
$\tau_1=\tau_2^*$. What is interesting (but not surprising, though) one
cannot have non-interacting fields passing to the real components.
Precisely, fields will be quadratically coupled in the Lagrangian. This
means that the usual intuition about field properties (like signs of
coefficients in front the kinetic term or the mass term) may not work.
In addition to the latter system of equations we can use the equations
of motion for the scalar fields which are
\begin{equation}
\label{FrEOMtaucsf}
\begin{split}
\ddot\tau_1+3H\dot\tau_1+J\tau_1=0, \quad
\ddot\tau_2+3H\dot\tau_2+J^*\tau_2=0
\end{split}
\end{equation}
Equations (\ref{FrEOMtauc}) and (\ref{FrEOMtaucsf}) describe the late
time evolution of the model with Lagrangian (\ref{action_model}). This
model possess a solution with the scalar field tending to the minimum
of the potential (i.e. $\tau\to0$) and the Hubble parameter going to
the constant. Such solution was constructed numerically and was proven
to be a solution \cite{Jukovskaya0707}. Also the asymptotic form of
this solution was derived in \cite{Koshelev07}. The idea is to compute
a solution to (\ref{FrEOMtaucsf}) using the constant $H=H_0$ and then
compute the correction to $H$ using (\ref{FrEOMtauc}). Then the
procedure can be iterated to give higher and higher corrections. It was
proven in \cite{Koshelev07} such iterations do converge.

Solution to (\ref{FrEOMtaucsf}) with constant $H=H_0$ is obviously
\begin{equation}
\label{FrEOMtaucsfsoltau}
\begin{split}
\tau_1&=\tau_{1+}e^{\alpha_+t}+\tau_{1-}e^{\alpha_-t},~
\tau_2=\tau_{2+}e^{\alpha_+^*t}+\tau_{2-}e^{\alpha_-^*t}\\
\end{split}
\end{equation}
where $\alpha_{\pm}=\frac{3H_0}{2}\left(-1\pm\sqrt{1-\frac{4J}{9H_0^2}}\right)$.
Consider $\tau_1$. Only one term in the solution converges when $t\to\infty$ in general (if both converge we select the slowest one). Let's assume it is the first one proportional to $\tau_{1+}$ constant. Then in order to pick the (slowest) convergent solution we put $\tau_{1-}=0$. The similar term with $\tau_{2+}$ will converge in $\tau_2$\footnote{Here we adopt the rule $\sqrt{z^*}=\sqrt{z}^*$ meaning that the phase of the complex number runs in the interval $[-\pi,\pi)$ and for $z=re^{i\varphi}$ the square root is $|\sqrt{r}|e^{i\varphi/2}$.}. To assure the initial field $\tau$ is real we have to impose $\tau_{2+}=\tau_{1+}^*$. Here we define $\tau_{1+}\equiv\tau_0$ and $\alpha_+\equiv\alpha$.

The first correction to the constant Hubble parameter in case only decaying modes in $\tau$ are present is
\begin{equation}
\label{FrEOMtaucsfsolH}
\begin{split}
H&=H_0+h=H_0+h_0\left(\tau_1^2+{\tau_1^*}^2\right)
\end{split}
\end{equation}
Constant $h_0$ is not independent and is related with $\tau_0$. We note
that $h$ is of order $\tau^2$. The last expression is a good
approximation for $H$ in the asymptotic regime when $h\ll H_0$.

For the energy and the pressure we find
\begin{equation}
\label{rhop10csol}
\begin{split}
\rho&=\frac12\left[\Fc'(J)\tau_1^2\left(\alpha^2
+J\right)+\Fc'({J^*}){\tau_1^*}^2\left({\alpha^*}^2
+{J^*}\right)\right]+g_o^2\Lambda,\\
p&=\frac12\left[\Fc'(J)\tau_1^2\left(\alpha^2
-J\right)+\Fc'({J^*}){\tau_1^*}^2\left({\alpha^*}^2
-{J^*}\right)\right]-g_o^2\Lambda
\end{split}
\end{equation}
and consequently for the state parameter and the speed of sound
\begin{equation}
\label{rhop10csolwcs}
\begin{split}
w&\approx-1-\frac{\Fc'(J)J\tau_1^2+\Fc'({J^*})J^*{\tau_1^*}^2}{g_o^2\Lambda},\\
c_s^2&=\frac{\Fc'(J)\alpha\tau_1^2\left(\alpha^2
-J\right)+\Fc'({J^*})\alpha^*{\tau_1^*}^2\left({\alpha^*}^2
-{J^*}\right)}{\Fc'(J)\alpha\tau_1^2\left(\alpha^2
+J\right)+\Fc'({J^*})\alpha^*{\tau_1^*}^2\left({\alpha^*}^2
+{J^*}\right)}.
\end{split}
\end{equation}
Further one can find the scale factor to be
\begin{equation}
\label{rhop10csola}
a=a_0\exp\left(H_0t+\frac{h_0}2\left(\frac{\tau_1^2}{\alpha}+\frac{{\tau_1^*}^2}{\alpha^*}\right)\right).
\end{equation}

It is manifest that the phantom divide is crossed thus periodically triggering the phantom phase of the evolution. The next problem is the behavior of the cosmological perturbations.
The main question: do they grow as one would expect in the ordinary phantom model or not?


\section{Cosmological perturbations in the neighborhood of the solution with complex masses}

Configurations with a single scalar field were widely studied and those appearing in
the non-local models do not have any distinguished properties. Roughly speaking configurations with many scalar fields were
explored as well but we have here new models featuring complex masses and complex coefficients in  front of the kinetic terms. As it was
stressed above there is no problem with this for the physics of our models while properties of such models, in particular the 
cosmological perturbations with such scalar fields were not studied in depth. Thus we focus on perturbations in the configuration with
complex roots $J$. The simplest case is one pair of complex conjugate roots where the background quantities were derived in
previous Section.



To find out the energy density perturbations we must use a system of equations on $\varepsilon$ which is the invariant energy density perturbation
and on $\zeta$ which is the gauge invariant scalar field perturbation (see, for instance, \cite{bardeen,Mukhanov,hwangnoh}). The system is 
\begin{equation}
\begin{split}
&(\varrho+p)\left(\ddot{\zeta}+\left(3H_0+\alpha+\alpha^*\right)\dot\zeta+\left(-3\dot H+\frac{k^2}{a_0^2}e^{-2H_0t}\right)\zeta\right)=\\
=&\left(\frac{J}{\alpha}-\frac{{J^*}}{\alpha^*}\right)
\left(\left[{\Fc'(J^*){\alpha^*}^2{\tau_1^*}^2}-{\Fc'(J)\alpha^2\tau_1^2}
\right]\dot\zeta+{2g_o^2\Lambda}\varepsilon\right)
\end{split}
\label{deltaijepsex}
\end{equation}
\begin{equation}
\begin{split}
&(\varrho+p)\left(\ddot\varepsilon+\dot\varepsilon H_0(8+3c_s^2)
+\varepsilon\left(15H_0^2+9H_0^2c_s^2+\frac{k^2}{a_0^2}e^{-2H_0t}\right)\right)=\\
=&\frac{2k^2\Fc'(J)\Fc'(J^*)\alpha^2{\alpha^*}^2\tau_0^2{\tau_0^*}^2}
{a_0^2g_o^2\Lambda}\left(\frac{J}{\alpha}
-\frac{{J^*}}{\alpha^*}\right){e^{2(-H_0+\alpha+\alpha^*)t}}\zeta.
\end{split}
\label{deltaepsijex}
\end{equation}
where we should use
\begin{equation*}
\begin{split}
\dot H&=2h_0\left(\tau_1^2\alpha+{\tau_1^*}^2\alpha^*\right)\\
\rho+p&=\Fc'(J)\tau_1^2\alpha^2+\Fc'({J^*}){\tau_1^*}^2{\alpha^*}^2\\
{c_s}^2&=\frac{\Fc'(J)\alpha\tau_1^2\left(\alpha^2
-J\right)+\Fc'({J^*})\alpha^*{\tau_1^*}^2\left({\alpha^*}^2
-{J^*}\right)}{\Fc'(J)\alpha\tau_1^2\left(\alpha^2
+J\right)+\Fc'({J^*})\alpha^*{\tau_1^*}^2\left({\alpha^*}^2
+{J^*}\right)}
\end{split}
\end{equation*}
Here $H$ is given by (\ref{FrEOMtaucsfsolH}), $a$ is given by (\ref{rhop10csola}), energy and pressure
are given by (\ref{rhop10csol}) and $w$ and $c_s^2$ follow from (\ref{rhop10csolwcs}). Note that $\varepsilon$ is a real function
and $\zeta$ is an imaginary function of time only.
The latter system of equations is ready to be solved numerically but in
order to get some insight in what is going on it is instructive to make
some assumptions about the value $J$. This makes some analytic progress
possible.

We recall the SFT origin of the model. Practically this means that
values of $J$ are determined with the string scales while $H_0$ is
expected to be much smaller. Therefore, it is natural to assume that
$|\sqrt{J}|\gg H_0$. This implies
\begin{equation*}
\begin{split}
\alpha\approx i\sqrt{J},&~\alpha^2\approx -J-i3H_0\sqrt{J}\\
\alpha^*\approx -i\sqrt{J^*},&~{\alpha^*}^2\approx -J^*+i3H_0\sqrt{J^*}
\end{split}
\end{equation*}
under this assumption the system (\ref{deltaijepsex},\ref{deltaepsijex}) becomes
\begin{equation}
\begin{split}
&(\varrho+p)\ddot{\zeta}+2i\left({\Fc'(J)J\tau_1^2\sqrt{J^*}
-\Fc'(J^*){J^*}{\tau_1^*}^2}\sqrt{J}\right)\dot\zeta+\\
+&\left(-i6h_0(\varrho+p)(\sqrt{J}\tau_1^2-\sqrt{J^*}{\tau_1^*}^2)+(\varrho+p)\frac{k^2}{a_0^2}\right)\zeta=\\
=&-i\left({\sqrt{J}}+{\sqrt{J^*}}\right){2g_o^2\Lambda}\varepsilon
\end{split}
\label{deltaijepsexM}
\end{equation}
\begin{equation}
\begin{split}
&(\varrho+p)\ddot\varepsilon+\dot\varepsilon 2i(\Fc'(J)J\sqrt{J}\tau_1^2-\Fc'({J^*})J^*\sqrt{J^*}{\tau_1^*}^2)-\\
-&\varepsilon\left(\left(\frac{k^2}{a_0^2}-6i
H_0\sqrt{J}\right)\Fc'(J)J\tau_1^2+\left(\frac{k^2}{a_0^2}+6iH_0
\sqrt{J^*}\right)\Fc'({J^*})J^*{\tau_1^*}^2\right)=
\\
=&-i\frac{2k^2\Fc'(J)\Fc'(J^*)J{J^*}}{a_0^2g_o^2\Lambda}\left(\sqrt{J}+\sqrt{{J^*}}\right)\tau_1^2{\tau_1^*}^2\zeta.
\end{split}
\label{deltaepsijexM}
\end{equation}
where we should use
\begin{equation*}
\begin{split}
\rho+p&=-(\Fc'(J)\tau_1^2J+\Fc'({J^*}){\tau_1^*}^2{J^*})
\end{split}
\end{equation*}

Using the explicit expression for $\tau_1=\tau_0e^{\alpha
t}\approx\tau_0e^{i\sqrt{J}t}$ and representing $\alpha=x/2+iy/2$ one
can write for any $A$
\begin{equation*}
\begin{split}
A\tau_1^2+A^*{\tau_1^*}^2&=2A_0e^{xt}\cos(yt+\varphi)\\
-i(A\tau_1^2+A^*{\tau_1^*}^2)&=2A_0e^{xt}\sin(yt+\varphi)\\
\end{split}
\end{equation*}
Introducing $\chi=i\frac{R}{g_o^2\Lambda}e^{xt}\zeta$ the equations of interest can be written as
\begin{equation}
\begin{split}
&\cos(yt+r)\ddot{\chi}+2(\sqrt{x^2+y^2}\sin(yt+r+\varphi_b)-x\cos(yt+r))\dot\chi+\\
+&\left(\cos(yt+r)\left(6h_0\sqrt{x^2+y^2}e^{xt}\sin(yt-\varphi_b)+\frac{k^2}{a_0^2}+x^2\right)-\right.\\
&\left.-2x\sqrt{x^2+y^2}\sin(yt+r+\varphi_b)\right)\chi=-{2y}\varepsilon
\end{split}
\label{deltaijepsexMcos}
\end{equation}
\begin{equation}
\begin{split}
&\cos(yt+r)\ddot\varepsilon+2\sqrt{x^2+y^2}\sin(yt+r-\varphi_b)\dot\varepsilon+\\
+&3H_0\sqrt{\left(\frac{k^2}{3a_0^2H_0}-x\right)^2+y^2}\cos(yt+r-\varphi_c)\varepsilon=\frac{2k^2y}{a_0^2}\chi.
\end{split}
\label{deltaepsijexMcos}
\end{equation}
where all the constant coefficients are real, $\varepsilon$ and $\chi$ are real, $x$ is expected to be negative, $R>0$, $r$ depends on $\Fc'(J)$, and
\begin{equation*}
\varphi_b=\arcsin\frac{x}{\sqrt{x^2+y^2}},~\varphi_c=\arcsin\frac{y}{\sqrt{(\frac{k^2}{3a_0^2H_0}-x)^2+y^2}}.
\end{equation*}

The most alarming points of the evolution are $yt+r=\frac{\pi}2+n\pi$ where the coefficients in front of second derivatives
become zero.
Numeric integration may hit problems at these points if the precision is not very high.
In the neighborhood of these points one has
\begin{equation}
\begin{split}
&t\ddot{\chi}-2\dot\chi+2x\chi=2\varepsilon
\end{split}
\label{deltaijepsexMcos0}
\end{equation}
\begin{equation}
\begin{split}
&t\ddot\varepsilon-2\dot\varepsilon-3H_0\varepsilon=\frac{2k^2}{a_0^2}\chi.
\end{split}
\label{deltaepsijexMcos0}
\end{equation}
For negative $x$ the solution for $\varepsilon$ around $t=0$ is $\varepsilon=\varepsilon_0+\varepsilon_1t+\dots$ meaning that these points are not singular
for the above system of equations.

A typical behavior for the function $\varepsilon$ is dumped oscillations.
Such a behavior does not depend on the wavenumber meaning that perturbations with complex conjugate scalar fields do vanish. This is different from
usual models with real scalar fields where different regimes exist and most likely growing modes are present.

This analysis can be easily extended to many complex roots and the main result, absence of growing modes at large times will remain.
Indeed, the background solution is determined with the component having largest $|x|$. Thus the background quantities can be considered
approximately the same. New equations (in total we have $N$ equations if we have $N$ roots) can also be accounted numerically and one can demonstrate
that the main effect of absence of the growing perturbations in $\varepsilon$ does remain. The best way to see this, however, is to go
to the exactly solvable system in which the analysis is more transparent. The detailed analysis of this question is to be found in the forthcoming paper
\cite{GK}.


\section{Coupling to dilaton and the cosmological signature of non-local models}

It is natural to consider a more general coupling of the open string modes with the closed string excitations.
The first non-trivial step is to add the dilaton to the metric.
The action (\ref{action_model_pre}) is minimally modified as follows 
\begin{equation}
S=\int d^4x\sqrt{-g}\left(\frac{e^{-\Phi}}{16\pi
G_N}(R+(\pd_\mu\Phi)^2-U(\Phi))+\frac1{g_o^2}\left(e^{-\Phi/2}\left(-\frac12\pd_\mu \tilde T\pd^\mu
\tilde T+\frac1{2}\tilde T^2\right)-\widetilde{e^{-\Phi/2}}{v(\bar{T})}\right)-\Lambda'\right). \label{action_model_d}
\end{equation}
Such a modified model was analyzed in details in \cite{AKd}. Here $\Phi$ is the dilaton field and $U(\Phi)$ is the dilaton potential.
The main point is that on one hand the dilaton field is dynamical while on the other
hand the restrictions on the rate of change of the Newtonian constant are very narrow (for example, from the Solar system measurements).
The latter circumstance allows to use a series expansion for the dilaton field around some constant value.
Effectively this means that equations derived in previous Sections get modified with small corrections. However,
the presence of the dynamical dilaton field dictates more restrictions on solutions and modifies the possible vacuum expectation values
for the field $T$ as well. This in turn alters the spectrum of roots $J_i$ making the system significantly different from the the
just tachyon dynamics.

There is one exceptional case for the above system when the tachyon potential has a minimum with both first and second derivatives vanishing.
Physically phis means that the potential is flat in the minimum. This corresponds to the real root $J=0$ if $v(\bar{T})\sim\bar{T}^4$.
In this particular case dilaton corrections render small deviations from $J=0$. These corrections
become comparable to the Hubble scale. It is possible because of the smallness of the dilaton rate of change.
As a consequence all the parameters in the solution turn out to be of the present cosmological scales.
The resulting behavior are oscillations with a period of order of 1~Gyr in the state parameter $w$ which is detectable.
Moreover, there are indications of oscillations of this kind in the literature \cite{V}. 

It is the situation which one cannot easily achieve
having just an open string sector while normally all the string quantities are of the order of $\AP$.


\section{Summary and outlook}

To summarize we have put in line the appearance of non-local models from SFT, construction of asymptotic solutions
in such models, analysis of perturbations and a possibility to extract observationally detectable effects.

Indeed, models coming out of the UV-complete theory are interesting because they are expected to be free of 
pathologies. Appearance of non-localities is natural in SFT while their impact on cosmological models is a novel effect.
Solutions in the non-local models in general are not constructed yet.
There are numeric attempts while analytic solutions are known only in the linear models.

The example of perturbations with complex roots carried out in this note demonstrates that linear perturbations can be confined thus
not destroying the system. It is not evident from the very beginning and supports the claim that the SFT based models are stable.
The case of complex $J_i$ has never
been studied in general and deserves deeper investigation. This problem and
other technical questions related to the equations of the present paper
will be addressed in the forthcoming publications.

Another interesting question is the study of models which possess exact solutions preserving the asymptotic behavior. This would drastically
simplify the analysis of the perturbations. Also generalization of the localization method to the presence of the dilaton field is
a yet unexplored problem worth further investigation.

As a more ambitious problem which is of great importance is a
construction of the formalism analogous to presented in this paper for
a model with self-interacting non-local scalar field. Such models play
important role in the SFT. For instance, rolling tachyon dynamics is
governed by action (\ref{action_model}) with a polynomial potential of
fourth degree. However, even background solutions are not very well
understood because there is no general analytic way of solving
non-local non-linear equations. On the other hand it follows from the
present analysis that passing to a local system with many fields is
vital for the construction of perturbation equations.

Looking a step further it is interesting to consider perturbations in
other non-local models coming from the SFT. For instance, models where
open and closed string modes are non-minimally coupled may be of
interest in cosmology. An example of the classical solution is
presented in \cite{AKd}. Furthermore it should be possible to extend
the formalism presented in this paper to other models involving
non-localities like modified gravity setups.


\section*{Acknowledgments}
The author would like to thank the organizers of the Invisible Universe 2009 meeting for the opportunity to present the work and for
creating the very stimulating environment for scientific discussions.
The author is grateful to I.Ya.~Aref'eva and S.Yu.~Vernov for the collaboration on papers led to this presentation and to
F.~Bezrukov, B.~Craps, B.~Dragovich, G.~Dvali, and
V.~Mu\-kha\-nov for useful comments and stimulating discussions. This work is
supported in part by RFBR grant 08-01-00798, state contract of
Russian Federal Agency for Science and Innovations 02.740.11.5057, the Belgian Federal Science Policy Office
through the Interuniversity Attraction Poles IAP VI/11, the European
Commission FP6 RTN programme MRTN-CT-2004-005104 and by FWO-Vlaanderen
through the project G.0428.06.


\end{document}